\documentclass[sigconf,nonacm]{acmart}
\graphicspath{ {images/} }
\usepackage{subcaption}
\usepackage{caption}

\AtBeginDocument{%
  \providecommand\BibTeX{{%
    \normalfont B\kern-0.5em{\scshape i\kern-0.25em b}\kern-0.8em\TeX}}}

\setcopyright{acmcopyright}
\copyrightyear{2018}
\acmYear{2018}
\acmDOI{XXXXXXX.XXXXXXX}

\acmConference[Conference acronym 'XX]{Make sure to enter the correct
  conference title from your rights confirmation emai}{June 03--05,
  2018}{Woodstock, NY}
\acmPrice{15.00}
\acmISBN{978-1-4503-XXXX-X/18/06}

\begin{document}

\title[Pedestrian-Vehicle Interaction in Shared Space]{Pedestrian-Vehicle Interaction in Shared Space: Insights for Autonomous Vehicles}

\author{Yiyuan Wang}
\email{yiyuan.wang@sydney.edu.au}
\affiliation{%
  \institution{Design Lab, Sydney School of
Architecture, Design and Planning, The University of Sydney}
  \city{Sydney}
  \country{Australia}
}

\author{Luke Hespanhol}
\email{luke.hespanhol@sydney.edu.au}
\affiliation{%
  \institution{Design Lab, Sydney School of
Architecture, Design and Planning, The University of Sydney}
  \city{Sydney}
  \country{Australia}
}

\author{Stewart Worrall}
\email{stewart.worrall@sydney.edu.au}
\affiliation{%
  \institution{Australian Centre for Field Robotics, \\
  The University of Sydney}
  \city{Sydney}
  \country{Australia}
}

\author{Martin Tomitsch}
\email{martin.tomitsch@sydney.edu.au}
\affiliation{%
  \institution{Design Lab, Sydney School of
Architecture, Design and Planning, The University of Sydney}
  \city{Sydney}
  \country{Australia}
}

\renewcommand{\shortauthors}{Wang et al.}

\begin{abstract}
Shared space reduces segregation between vehicles and pedestrians and encourages them to share roads without imposed traffic rules. The behaviour of road users (RUs) is then controlled by social norms, and interactions are more versatile than on traditional roads. Autonomous vehicles (AVs) will need to adapt to these norms to become socially acceptable RUs in shared spaces. However, to date, there is not much research into pedestrian-vehicle interaction in shared-space environments, and prior efforts have predominantly focused on traditional roads and crossing scenarios. We present a video observation investigating pedestrian reactions to a small, automation-capable vehicle driven manually in shared spaces based on a long-term naturalistic driving dataset. We report various pedestrian reactions (from movement adjustment to prosocial behaviour) and situations pertinent to shared spaces at this early stage. Insights drawn can serve as a foundation to support future AVs navigating shared spaces, especially those with a high pedestrian focus.
\end{abstract}




\maketitle

\section{Introduction}
The advancement of sensing and computing technologies enables autonomous vehicles (AVs) to increasingly follow traffic rules and operate safely on structured, regulated roads. However, in less predictive situations such as unmarked crossings and vehicle-pedestrian shared zones, pedestrians and drivers in traditional vehicles often rely on informal rules (e.g. movement, eye contact) to convey intentions and anticipate those from others \cite{predhumeau2021pedestrian,rasouli2019autonomous}. These informal rules, also referred to as social norms \cite{adrian2019glossary,rasouli2019autonomous}, are linked strongly with pedestrians' expectations \cite{schieben2019designing,predhumeau2021pedestrian} and can influence how acceptable a behaviour is in a particular traffic context \cite{adrian2019glossary,rasouli2019autonomous}. This leads to the question of equipping AVs with the ability to practice social norms in order to navigate more ambiguous and complex environments in good time and manner. In recent years, researchers have started to tackle this gap by learning driver-pedestrian interaction patterns \cite{rasouli2017agreeing,dey2017pedestrian} and designing external human-machine interfaces (eHMIs) for AVs to communicate with nearby pedestrians \cite{dey2020taming}. However, these studies in general have focused on crossing situations on normal roads, while a shared space environment where more diverse social interactions unfold among traffic participants is far less explored.

Shared space is a recent urban planning approach designed to minimise the demarcation between pedestrians and vehicles by removing kerbs, road markings, traffic signs, and traffic signals and therefore encourage all road users (RUs) to share the same public space with equal rights \cite{moody2014shared}. Besides reducing automobile dominance and improving pedestrian priority \cite{moody2014shared}, this approach aims further to increase pedestrian social activities and dwell time, promoting a lively, multi-functional street environment where a transport corridor also serves as a destination \cite{karndacharuk2013analysis}. The behaviour of RUs is then guided by social interactions \cite{hammond2013attitudes}, and pedestrians and cars will need to cooperate in diverse scenarios \cite{predhumeau2021pedestrian,li2021autonomous}. Yet, both pedestrians and drivers have reported feeling uneasy in shared spaces \cite{moody2014shared,kaparias2012analysing,hammond2013attitudes} and misunderstanding of intentions between them has been raised as a problem \cite{li2021autonomous}.

As shared spaces become more popular in urban planning \cite{clarke2006shared}, AVs will soon need to navigate crowds and encounter interactions that used to be handled by human drivers. Researchers and engineers have been continuously improving algorithms to help AVs plan paths and movement in shared spaces to respect social norms related to proxemics and kinematics \cite{chen2017socially,li2020socially}. However, the absence of drivers still makes it questionable whether AVs can deal with the complex surroundings and follow suitable interaction strategies in different situations. Moreover, there is a general lack of knowledge around real-life pedestrian-vehicle interaction in shared spaces or evidence supporting pertinent external interaction designs. It remains unclear what AVs should anticipate and what factors they should consider interacting with pedestrians in this context.

In this paper, we present a video observation investigating pedestrian reactions towards an automation-capable electric vehicle (EV) driven manually in multiple shared spaces across 14 months. The videos were recorded via vehicle-mounted cameras during naturalistic driving and capture scenes around the vehicle's vicinity. The fixed driving route covers two outdoor shared pedestrian areas and one underground car park. We identify a range of social responses from pedestrians ranging from movement adjustment to prosocial behaviour \cite{harris2014prosocial} as well as various critical situations that have safety or efficiency considerations.

The contribution of this paper is twofold. First, we observe naturally occurring pedestrian-vehicle interaction in shared spaces from a long-term video dataset. We report a variety of pedestrian behaviours towards the vehicle and depict important situations. Second, based on our observation, we provide insights contributing to external interaction designs for future AVs in shared spaces at this early, exploratory stage, especially for shared spaces focusing on pedestrian priority and activities.

\section{Related Work}
\subsection{Field Observations Informing AV-pedestrian Interaction}
Communication strategies between pedestrians and drivers have been studied since the introduction of automobiles in the last century \cite{lee2021road}. In recent years, many field observations have been conducted to understand current pedestrian-driver/vehicle interaction patterns for the purpose of informing interaction designs applicable to future AVs. Although these studies predominantly investigated crossing situations on regular roads, they have confirmed the important role that social norms play. For example, mutual attention via eye gaze is found to be a prominent signal for crossing and yielding intentions \cite{rasouli2017agreeing,nathanael2018naturalistic}. Pedestrians also engage in other non-verbal communications such as nodding and hand gestures to make their intentions more explicit when misunderstandings happen \cite{nathanael2018naturalistic} or to show gratitude and acknowledgement to the driver \cite{rasouli2017agreeing}. Some studies investigated the power of vehicle movement in communication \cite{risto2017human}. For instance, drivers' intent can be signalled by vehicles' stopping behaviours \cite{domeyer2019proxemics}, and vehicle speed profiles can influence pedestrian crossing decisions \cite{schneemann2016analyzing}. Besides, many studies discovered that movement, both vehicle and pedestrian ones, is the main communication channel compared to those more explicit cues (e.g. gestures) \cite{dey2017pedestrian,lee2021road,risto2017human,moore2019case}. However, this could be influenced by road types as pedestrians were found to establish more eye contact with drivers on lower-speed roads \cite{schneemann2016analyzing}, and it has been speculated that more explicit communication could occur in shared spaces \cite{lee2021road}.

More interesting findings regarding pedestrian reactions have been discovered when there was no perceivable driver in the vehicle. Some studies adopted a ``ghost driver'' approach where the human driver controlling the vehicle was hidden inside the driver's seat \cite{currano2018vamos,li2020road,moore2020defense}. Pedestrians exhibited curiosity, such as testing the vehicle and taking photos \cite{currano2018vamos,de2019perceived}. They also showed different crossing and looking duration when crossing in a group or alone \cite{li2020road}. Nonetheless, one study found pedestrians avoided the driverless vehicle \cite{de2019perceived}, and another study reported vandalism behaviours towards the vehicle, for example, blocking the vehicle's way deliberately or using abusive words \cite{moore2020defense}.

With the removal of traffic rules, pedestrians and vehicles will engage in more social interactions in shared spaces \cite{hammond2013attitudes}. One study recently observed pedestrian-driver looking behaviours and vehicle speed changes in a UK car park \cite{uttley2020road}. However, to our knowledge, there is no systematic investigation of a broader range of pedestrian behaviours and more diverse situations, which is the goal of our observation.

\subsection{Pedestrian Experiences in Shared Space}
In the early 2010s, a number of shared spaces projects were deployed to improve pedestrian priority in the urban environment. A study in the city centre of Auckland, New Zealand reported a higher pedestrian volume and increased pedestrian activities and dwell time after the shared space implementation \cite{karndacharuk2013analysis}. Nevertheless, studies in the UK revealed concerns from shared space users. Pedestrians still sought to use pedestrian-only facilities (e.g. dedicated crosswalks) \cite{moody2014shared,kaparias2012analysing} even though sometimes the location of those facilities lengthened their routes \cite{moody2014shared}. Other studies found pedestrians to report feeling anxious around high-speed motor vehicles and during peak traffic hours \cite{moody2014shared} and suggesting that they would feel safer when their presence was clear to other RUs (low vehicular speed, high pedestrian volume, and good lightening) \cite{kaparias2012analysing}. Additionally, issues related to wider RU groups were raised. Drivers indicated feeling uneasy around children and elderly \cite{kaparias2012analysing}, and pedestrians with impaired vision or mobility expressed concerns about their road priority \cite{hammond2013attitudes}.

More recently, pedestrian experiences have been gathered from deployments of fully autonomous shuttles in Europe. These shuttles operated driverlessly in mixed-road settings, and many came across vehicle-pedestrian shared zones. In the CityMobil2 project \cite{schieben2019designing,merat2018externally,madigan2019understanding}, pedestrians considered themselves to have higher priority over the shuttles in the absence of road markings \cite{merat2018externally} and suggested the need for the shuttles to display information related to manoeuvres and the awareness of other RUs \cite{schieben2019designing,merat2018externally}. One study highlighted the importance of ensuring the behaviour of AVs matches with RUs' expectations to avoid frustration and increase safety \cite{madigan2019understanding}. Likewise, pedestrians in the EasyMiles project expected the AVs to follow the behaviour of manually-driven vehicles and suggested simple forms of communication like horns and indicators \cite{locken2019investigating}. Moreover, pedestrians and cyclists in the WePods project felt significantly safer when sharing roads with WePods due to the slower speed than traditional vehicles \cite{rodriguez2017safety}. However, local residents reported staying away from the Sion SmartShuttle and also suggested the need for external information displays \cite{eden2017road}. Compared to these studies, we provide a more focused investigation that specifically analyses pedestrian-vehicle interaction in shared spaces and aims to inform external interaction designs for future AVs in this context.

\begin{figure*}[t]
    \centering
    \begin{subfigure}[b]{0.246\textwidth}
        \centering
        \includegraphics[width=\textwidth]{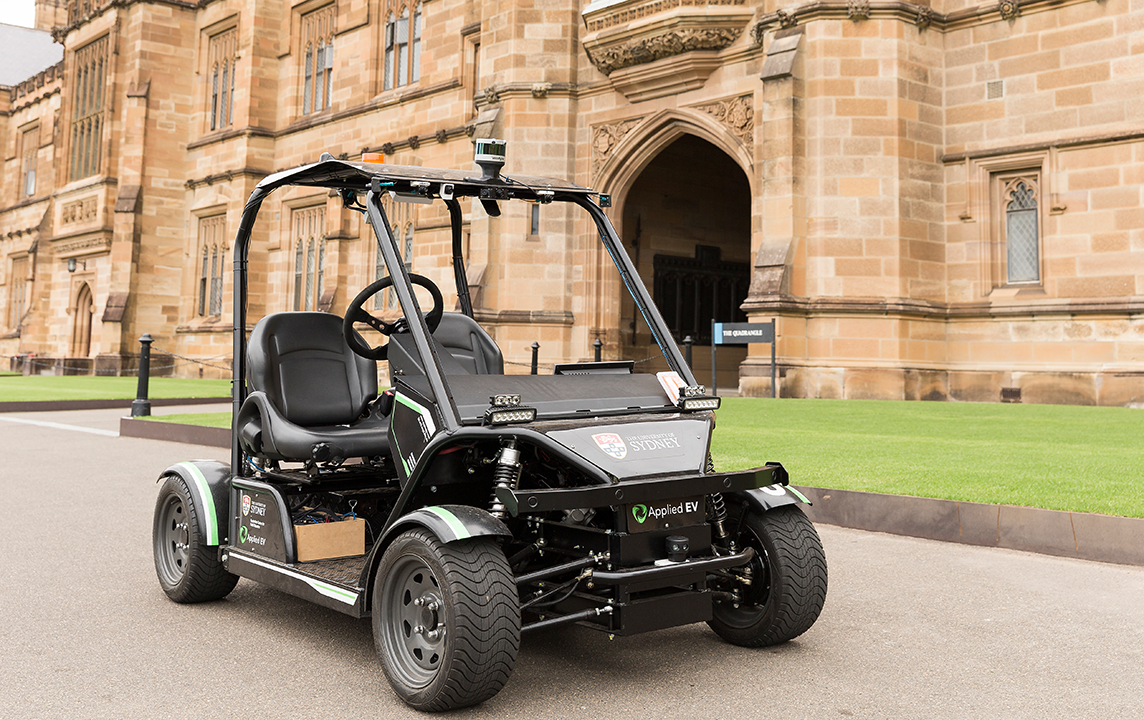}
        \caption{EV used for data collection}
        \label{vehicle}
    \end{subfigure}
    \begin{subfigure}[b]{0.246\textwidth}
        \centering
        \includegraphics[width=\textwidth]{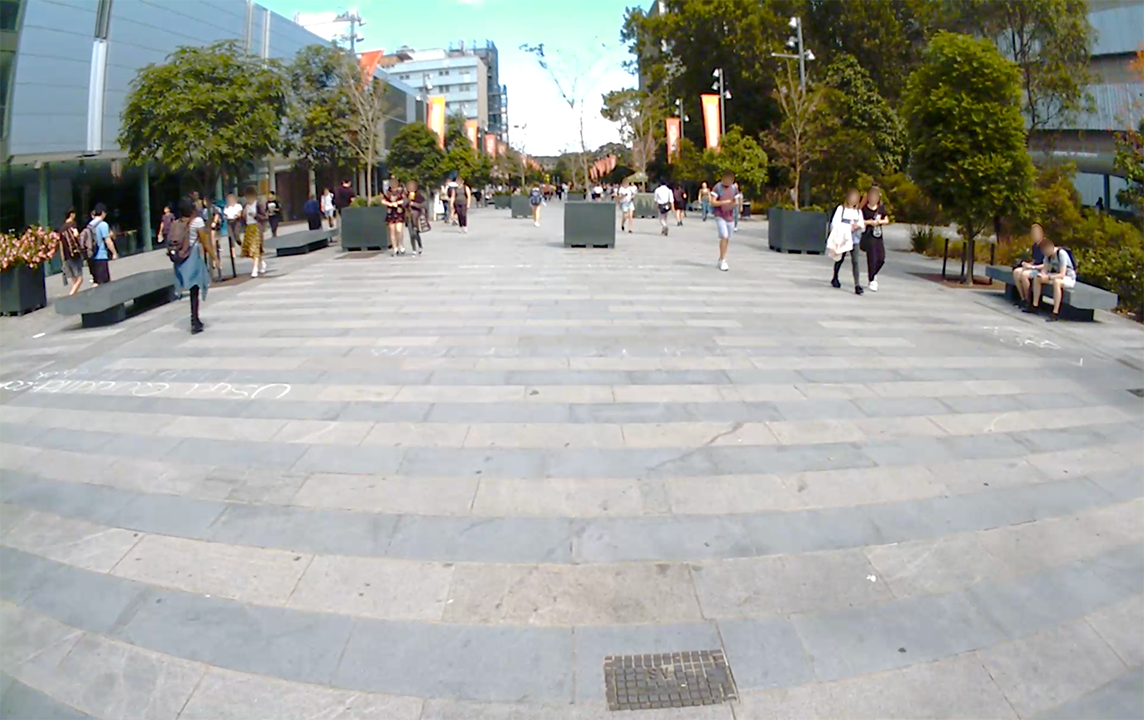}
        \caption{Eastern Avenue}
        \label{eastern_avenue}
    \end{subfigure}
    \begin{subfigure}[b]{0.246\textwidth}
        \centering
        \includegraphics[width=\textwidth]{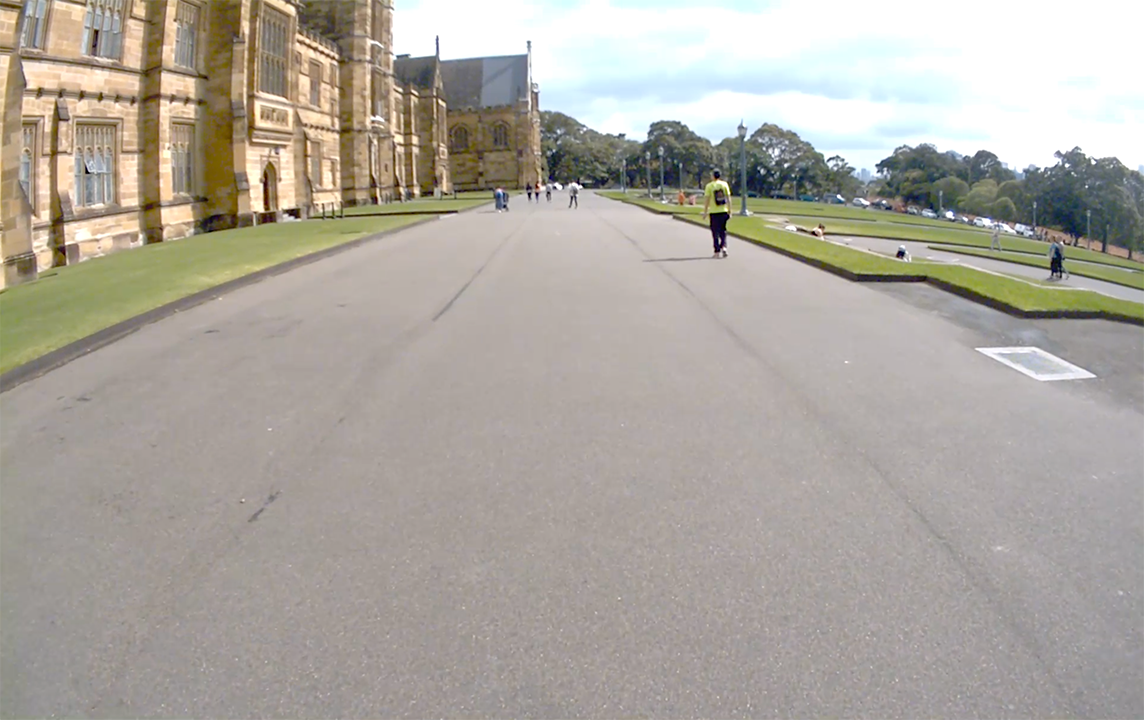}
        \caption{Quadrangle}
        \label{quadrangle}
    \end{subfigure}
    \begin{subfigure}[b]{0.246\textwidth}
        \centering
        \includegraphics[width=\textwidth]{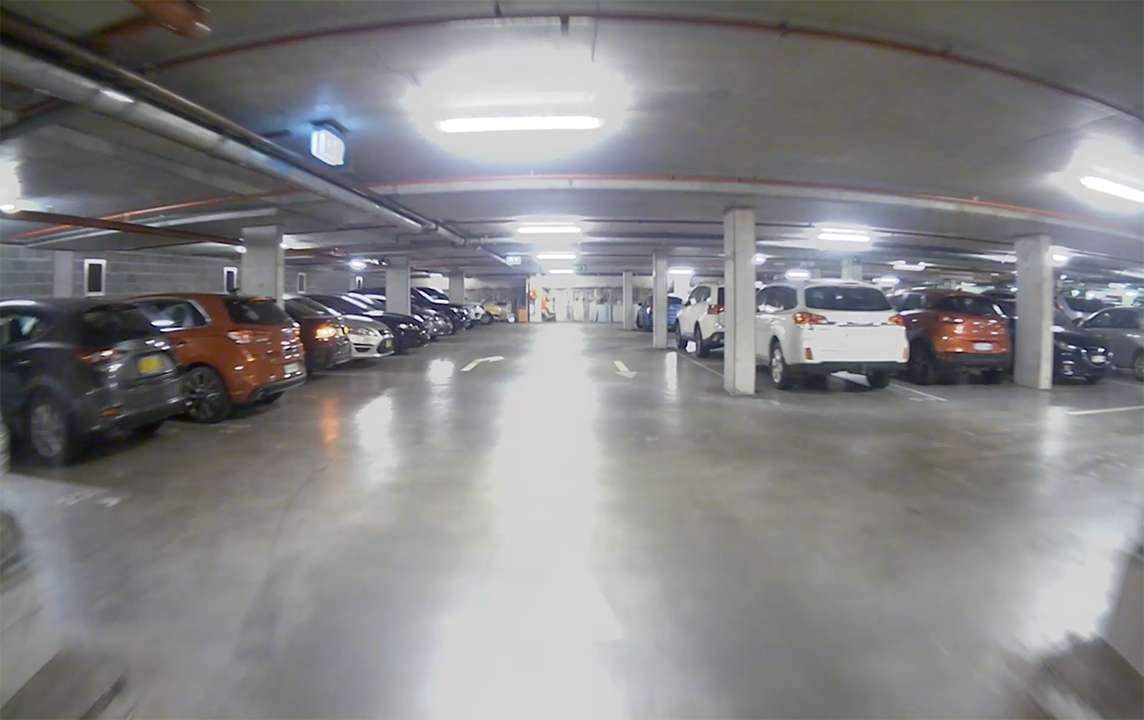}
        \caption{Underground car park}
        \label{car_park}
    \end{subfigure}
    \caption{Apparatus and environment for data collection.}
\end{figure*}

\subsection{AVs: Social Actors in Future Traffic}
Beyond technical aspects that ensure the safety of AVs, social acceptance related to human factors plays a critical role in the uptake of AVs in everyday traffic \cite{prattico2021comparing}. Public scepticism remains about AVs' ability to take actions intelligently like humans. When encountering an AV, pedestrians can avoid the vehicle \cite{de2019perceived} or walk ``extremely carefully'' around it \cite{eden2017road}. Moreover, field trials revealed hostile behaviours from pedestrians and other drivers \cite{moore2020defense}, such as forcing AVs into submission\footnote{https://www.theguardian.com/technology/2016/oct/30/volvo-self-driving-car-autonomous} or even attacking them with rocks\footnote{https://www.nbcnews.com/tech/innovation/humans-harass-attack-self-driving-waymo-cars-n950971}.

A promising approach to increasing AVs' social acceptance is the design of eHMIs \cite{prattico2021comparing}. They are intended as a replacement for driver cues and convey AVs' intent and awareness to nearby RUs \cite{dey2020taming}. Current eHMIs are mostly limited to the rational principle of collision avoidance \cite{schieben2019designing}, and only a few have carried social capacities more than passage negotiation. One study has presented an eHMI that conveys courtesy to pedestrians by displaying ``thank you'' and ``you're welcome'' in crossing scenarios \cite{colley2021investigating}. Another study has investigated AVs' prosocial communication (i.e. behaviours that benefit others) by displaying additional light cues for more vulnerable pedestrians like children and elderly \cite{sadeghian2020exploration}. Furthermore, it is found that ascribing human traits to AVs (e.g. name, gender, voice) increases people's trust \cite{waytz2014mind}. This is indeed a common approach to design likable and socially accepted robots in human-robot interaction (HRI) research \cite{bartneck2008measuring,wang2021can}. Some research has considered future AVs as a hybrid of conventional cars and mobile robots \cite{wang2021can,predhumeau2021pedestrian}, and hence it can be anticipated that future AVs can adopt communication skills similar to social robots.

Road users make decisions not only by complying with traffic rules but also by interpreting the ongoing traffic activities involving human behaviour, emotion, and other contextual factors \cite{eden2017road}. This suggests that for AVs to blend into future traffic, their external interaction should allow them to behave in a way compatible with the evolving surroundings. Thus, we seek to support AVs' role as social participants in traffic, especially in a non-traffic rule-based environment where social norms are constantly used.

\section{Method}
In this study, we used an existing video dataset, the USyd Campus Dataset\footnote{https://ieee-dataport.org/open-access/usyd-campus-dataset}, to observe pedestrian behaviours in response to an automation-capable vehicle driven manually in shared spaces. The videos were analysed through a combination of single open coding by the lead author and two workshops involving all authors to review and discuss the codes.

\subsection{Research Questions}
This study aims to answer two research questions: (RQ1) What are current pedestrian-vehicle interactions in shared spaces? (RQ2) What are insights gained for future designs around AV-pedestrian interaction in a shared space context?

\subsection{Data}
The USyd Campus Dataset was collected between March 2018 and April 2019 by the third author and their fellow researchers and published in IEEE Intelligent Transportation Systems Magazine \cite{zhou2020developing}. The videos were collected with a small, automation-capable EV (Fig.~\ref{vehicle}) driven by a human operator. A weekly drive was taken on and near The University of Sydney campus across a 14-month period, following a fixed route that covered regular roads and three shared spaces (two shared pedestrian areas and one underground car park) (Fig.~\ref{eastern_avenue}, \ref{quadrangle}, and \ref{car_park}). As no drive was taken during some weeks, the dataset contains a total of 52 drives, i.e. 52 weeks of data. Six NVIDIA gigabit multimedia serial link (GMSL) cameras were mounted on the vehicle to capture six different perspectives (front-middle, front-left, front-right, side-left, side-right, and rear), whilst the first 23 weeks only contain the three front-facing perspectives. For the purpose of this study, we only analysed videos recorded in the three shared spaces, which comprise approximately 5.6 hours of driving. The vehicle was driven naturally and consistently on the shared spaces with a slow speed around 7.6 km/h (2 m/s), maintained appropriate distance with nearby pedestrians by changing speed and steering, and did not stop unless being blocked by pedestrians in front. The operator minimised the use of the horn or direct communication with pedestrians to keep the driving less obtrusive to the environment.

\begin{figure*}[t]
  \centering
  \includegraphics[width=0.9\linewidth]{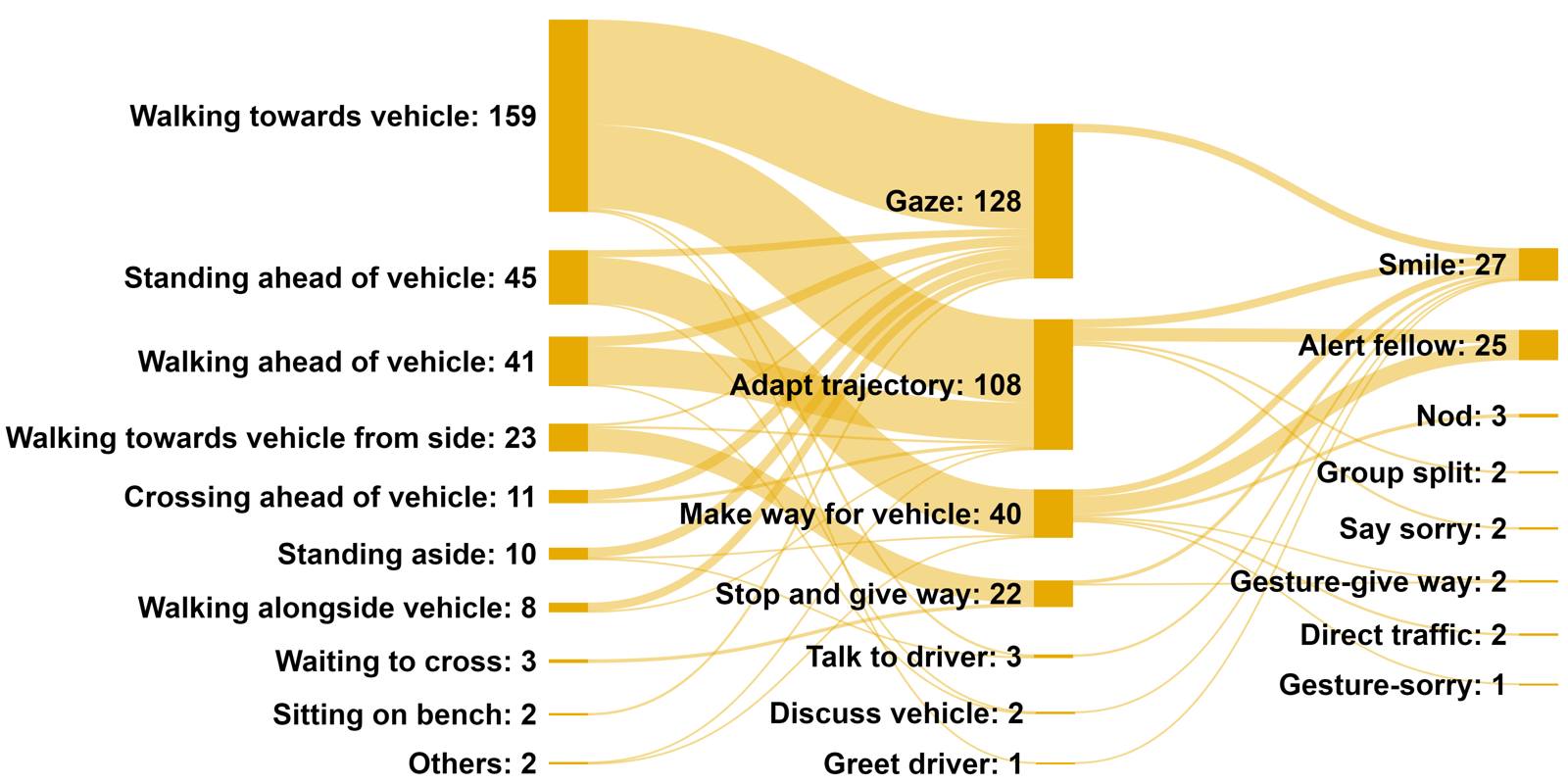}
  \caption{A summary of behavioural codes. Left column: pedestrian states (mutually exclusive); middle column: pedestrian behaviours (mutually exclusive); right column: pedestrian behaviours accompanying the behaviours in the middle column. Both sum of left column and sum of middle column equal the total number of interactions.}
  \label{sequence}
\end{figure*}

\subsection{Procedure}
The observation was conducted in the BORIS software \cite{friard2016boris} which synchronised the different perspectives and allowed for event coding. Spreadsheets were used as a supplementary coding tool to provide more descriptions of the interactions, with timestamps matched with the codes in BORIS. We excluded the side-left, side-right, and rear perspectives in order to (1) support data consistency since the first 23 weeks do not contain these three perspectives and (2) make the manual viewing process feasible as we needed to watch multiple perspectives simultaneously. Additionally, the lens of the cameras had a 100$^{\circ}$ horizontal field of view (FOV), and hence the three front perspectives were wide enough to capture pedestrians located on the sides; for example, when the vehicle was waiting as the lead car at a signalised crossing on a two-way, three-lane road, the front cameras were able to capture pedestrians waiting on both sides of the crossing.

The coding procedure was carried out in three steps. First, the lead author performed an inductive coding for the first half of the videos (26 weeks) and repeated the process on the same videos again to keep the codes consistent due to the inductive nature. A workshop was held among all authors (3 interaction designers, 1 engineer) to discuss the codes and watch representative scenes from the videos corresponding to the codes. In the second step, the lead author analysed the second half of the videos (26 weeks) using improved codes. All authors held a second workshop to review changes to the codes along with corresponding scenes and then reached an agreement on the final codes. In the third step, the lead author revisited and applied the final codes to all videos.

We coded interactions between pedestrians and the vehicle across the three shared spaces featured in the dataset (Fig.~\ref{eastern_avenue}, \ref{quadrangle}, and \ref{car_park}). Each interaction was observed from a single pedestrian or multiple pedestrians reacting as a unity towards the vehicle. An interaction was coded when the pedestrian(s) reacted to the vehicle, including changes in movement, facial or bodily expressions, and other responses associated with the vehicle (e.g. directed traffic for the vehicle to pass). We coded \textit{pedestrian behaviour} -- pedestrian reactions towards the vehicle, \textit{pedestrian state} -- the original state that they engaged in when the interaction happened, and \textit{pedestrian group size} -- whether they were in a group or alone. It should be noted that pedestrians could be labelled as in group even if they were the only one in that group reacting to the vehicle. We also recorded additional textual descriptions to capture notable interactions, e.g. descriptions of a risky situation.

\section{Results}
We recorded a total of 304 interactions, of which 249 (82\%) happened on Eastern Avenue, 49 at the Quadrangle (16\%), and 6 (2\%) in the underground car park. Fig.~\ref{distribution} shows the number of interactions recorded each week. In general, more interactions were observed during university semesters compared to holidays and mid-term breaks.

\subsection{Pedestrian Behaviours}
\begin{figure*}[t]
    \centering
    \begin{subfigure}[b]{0.31\textwidth}
        \centering
        \includegraphics[width=\textwidth]{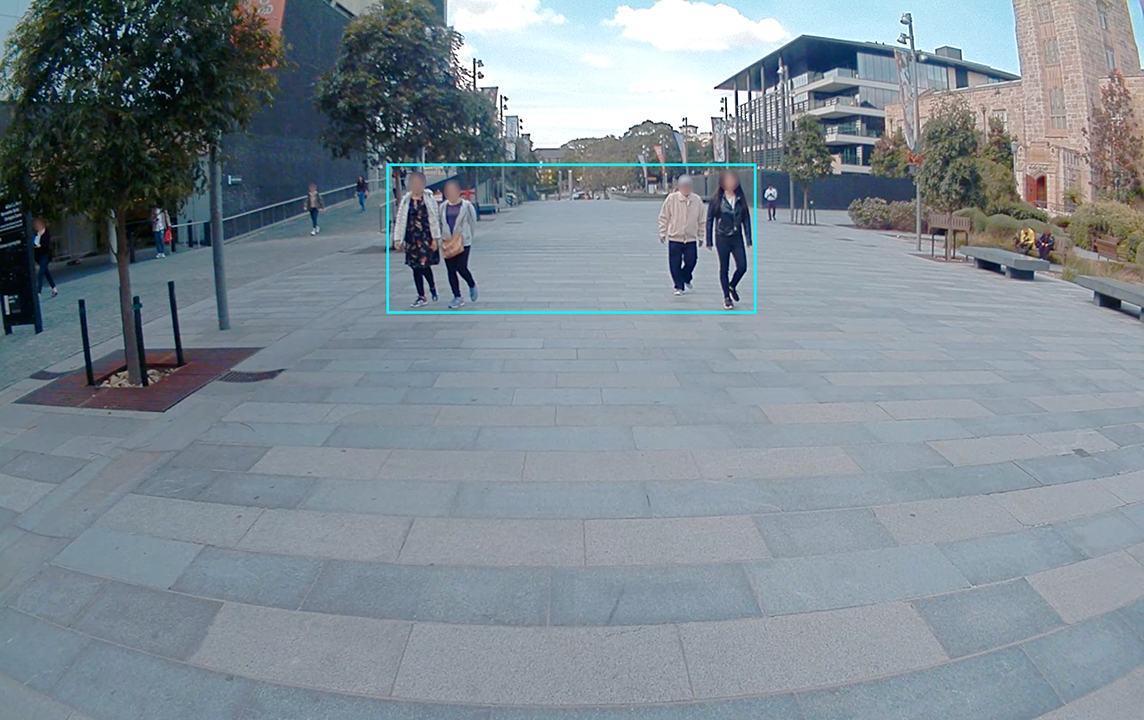}
        \caption{Walking towards vehicle}
        \label{walking_towards}
    \end{subfigure}
    \begin{subfigure}[b]{0.31\textwidth}
        \centering
        \includegraphics[width=\textwidth]{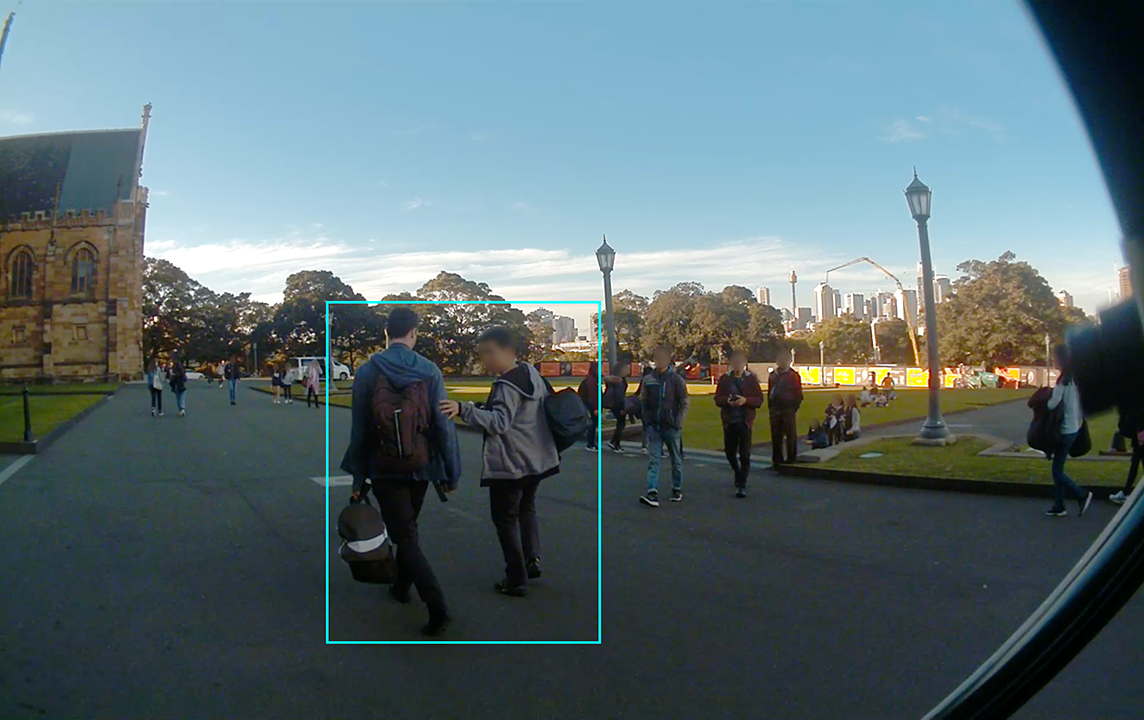}
        \caption{Walking ahead of vehicle}
        \label{walking_ahead}
    \end{subfigure}
    \begin{subfigure}[b]{0.31\textwidth}
        \centering
        \includegraphics[width=\textwidth]{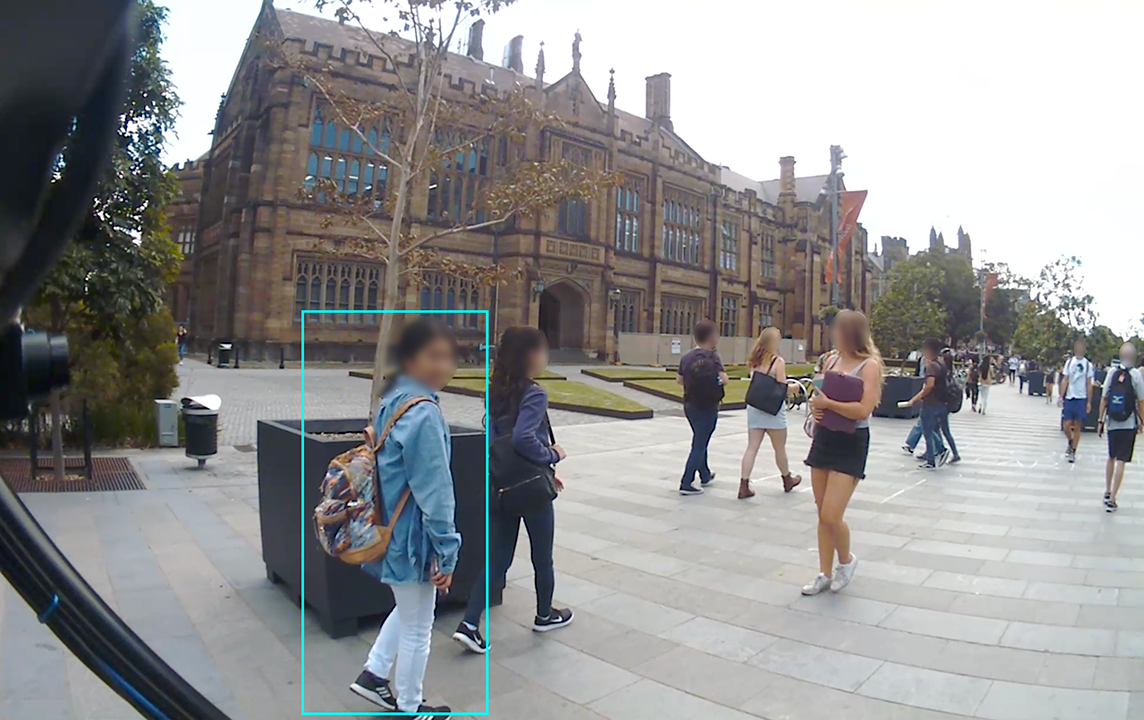}
        \caption{Walking alongside vehicle}
        \label{walking_alongside}
    \end{subfigure}
    \par\medskip
    \begin{subfigure}[b]{0.31\textwidth}
        \centering
        \includegraphics[width=\textwidth]{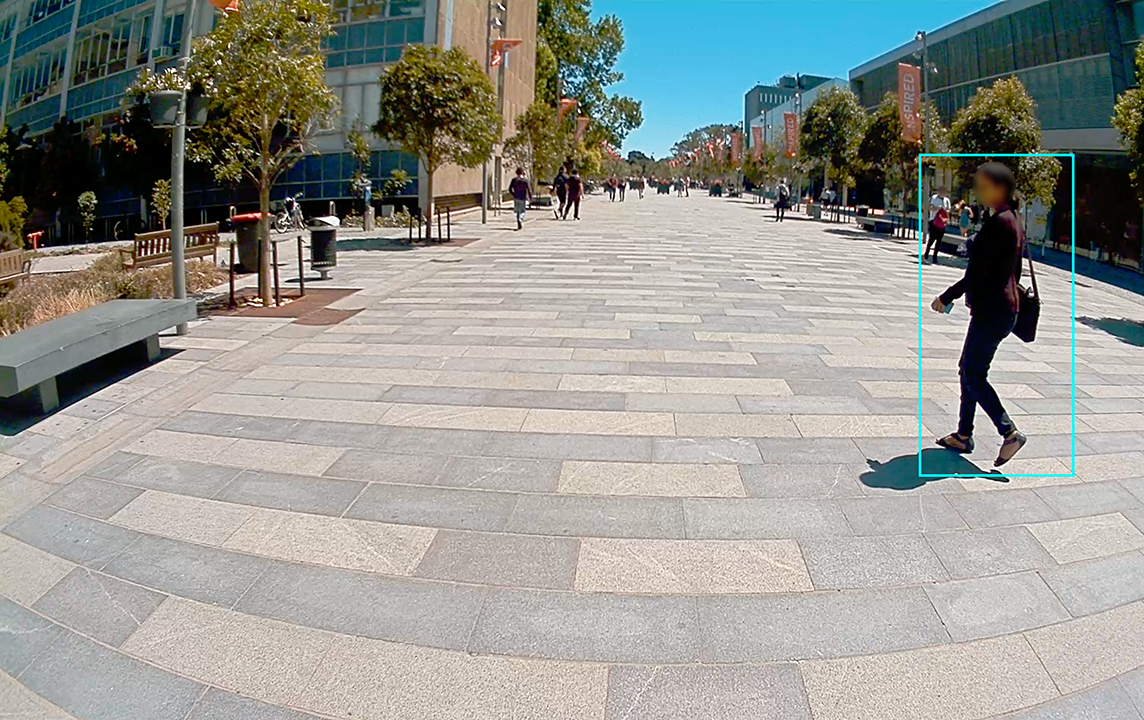}
        \caption{Walking towards vehicle from side}
        \label{walkingtowards_side}
    \end{subfigure}
    \begin{subfigure}[b]{0.31\textwidth}
        \centering
        \includegraphics[width=\textwidth]{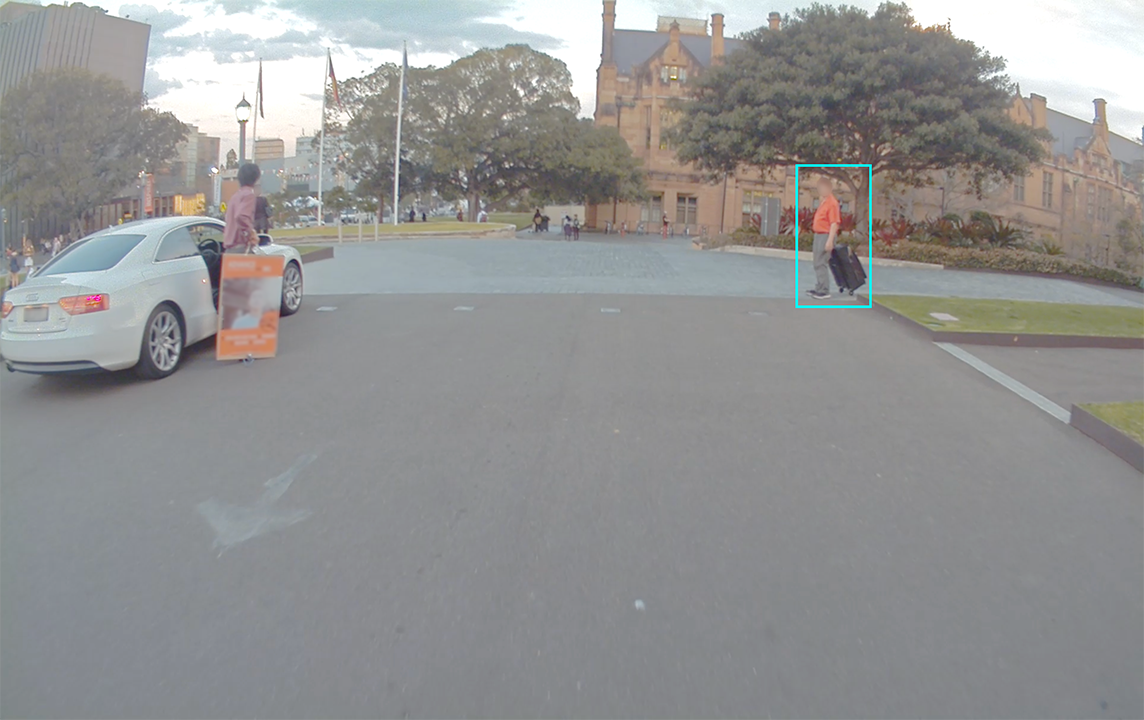}
        \caption{Waiting to cross}
        \label{waiting_cross}
    \end{subfigure}
    \begin{subfigure}[b]{0.31\textwidth}
        \centering
        \includegraphics[width=\textwidth]{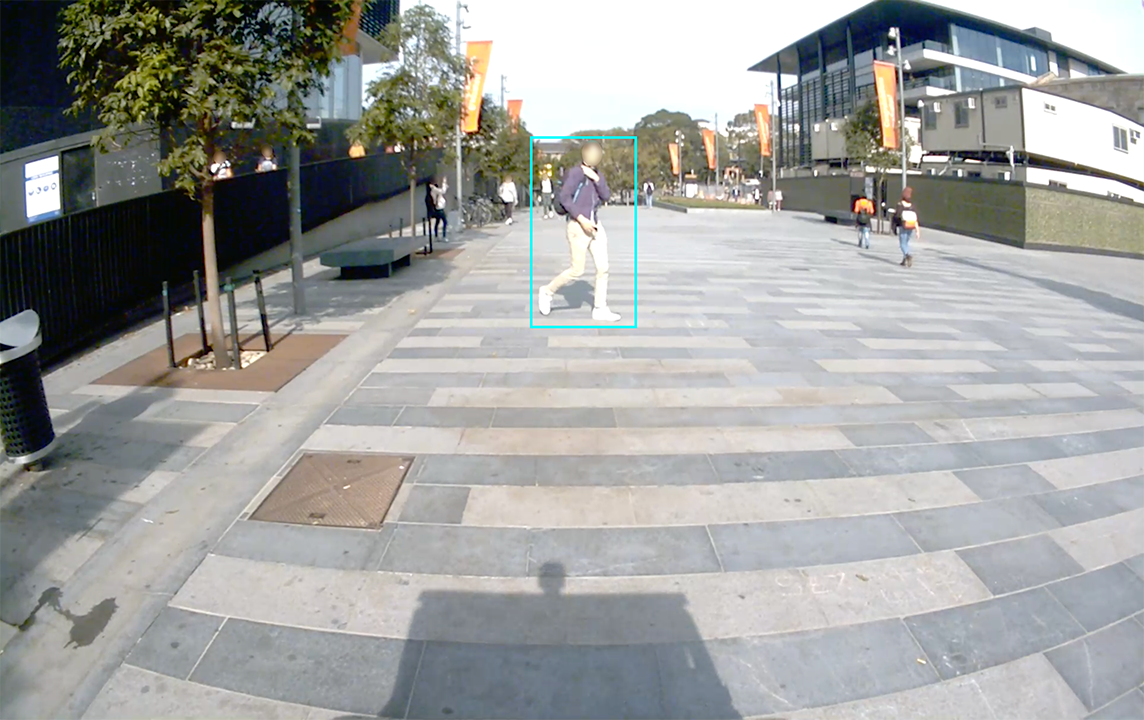}
        \caption{Crossing ahead of vehicle}
        \label{crossing}
    \end{subfigure}
    \par\medskip
    \begin{subfigure}[b]{0.31\textwidth}
        \centering
        \includegraphics[width=\textwidth]{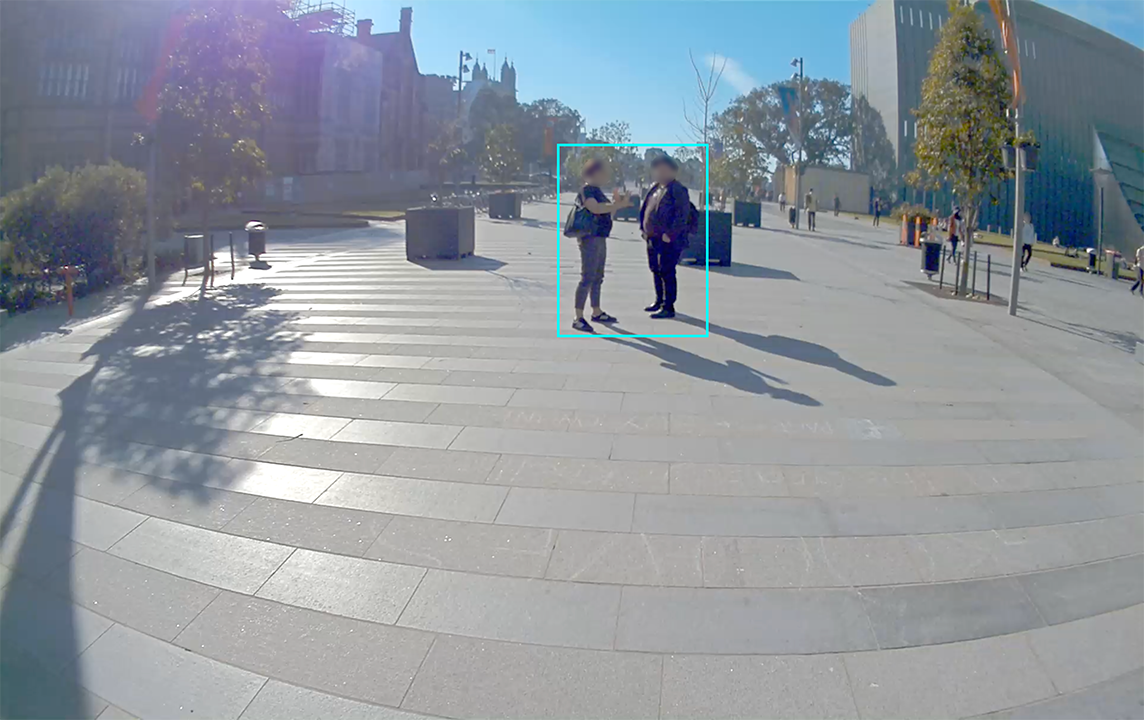}
        \caption{Standing ahead of vehicle}
        \label{standing_ahead}
    \end{subfigure}
    \begin{subfigure}[b]{0.31\textwidth}
        \centering
        \includegraphics[width=\textwidth]{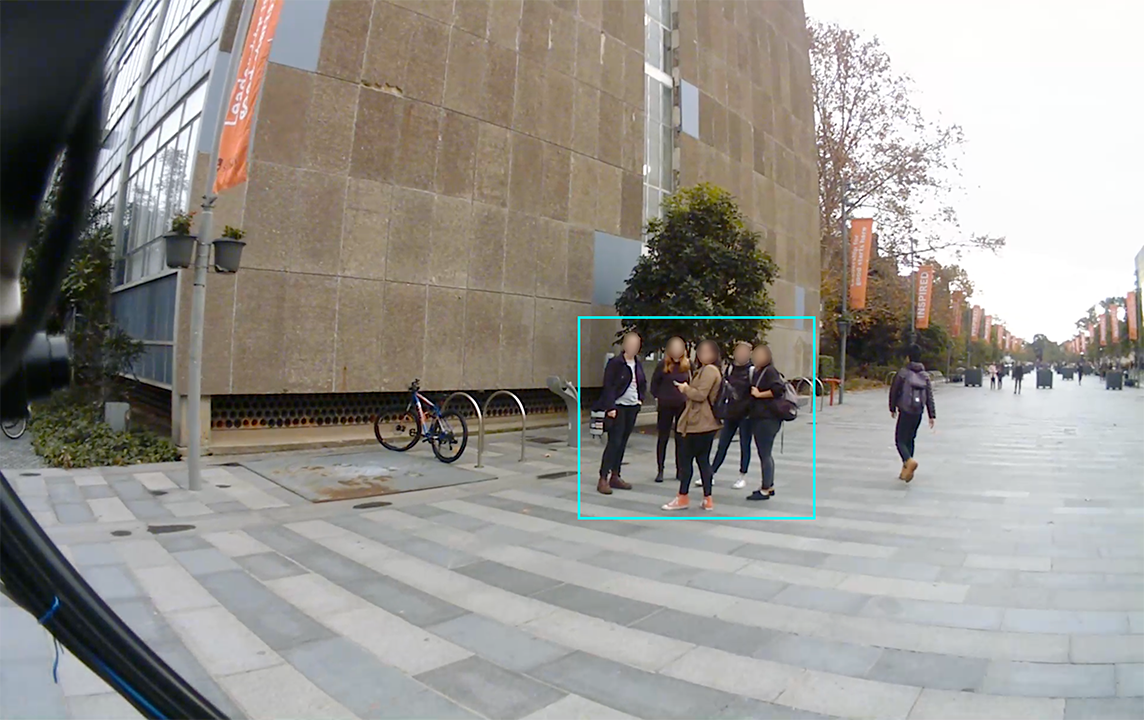}
        \caption{Standing aside}
        \label{standing_aside}
    \end{subfigure}
    \begin{subfigure}[b]{0.31\textwidth}
        \centering
        \includegraphics[width=\textwidth]{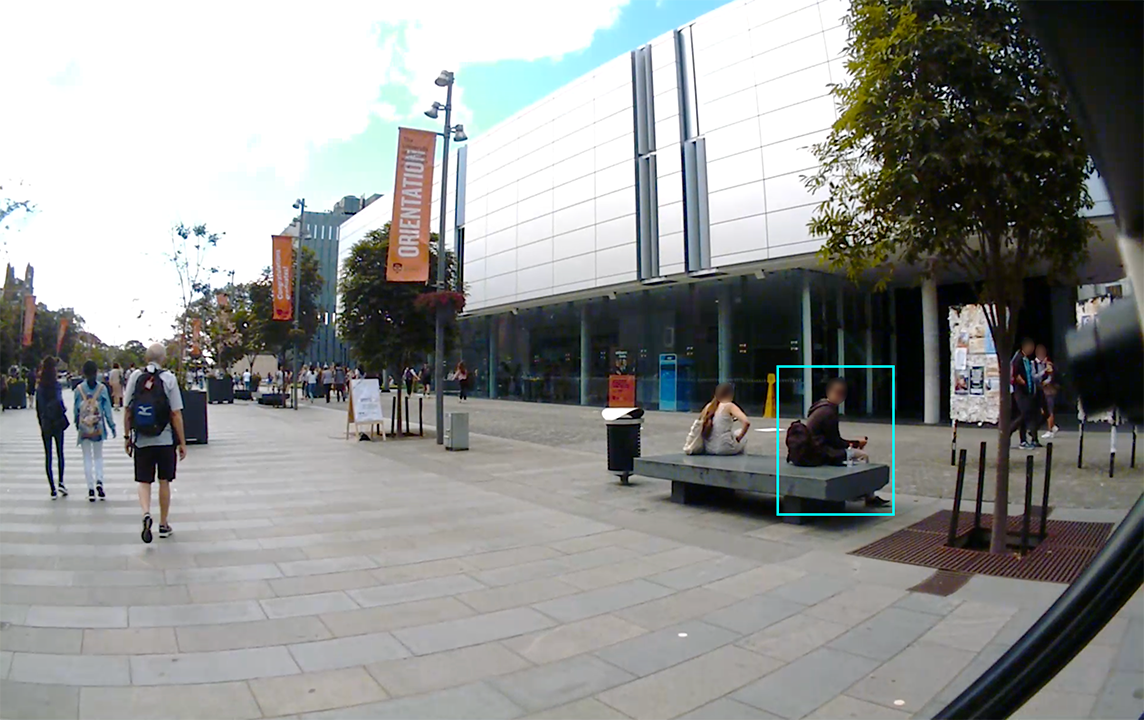}
        \caption{Sitting on bench}
        \label{sitting}
    \end{subfigure}
    \caption{Pedestrian states as shown in the sub-captions. We use cyan bounding boxes to highlight pedestrians referred to. More descriptions: (a) the group of pedestrians adapted trajectory and split into two groups; (b) one pedestrian was alerting the other pedestrian; (c) the pedestrian moved to the side; (d) the pedestrian saw the vehicle and stopped walking; (e,f,i) the pedestrians were gazing at the vehicle; (g) the pedestrians were chatting; (h) the group of pedestrians were gazing and smiling.}
    \label{state}
\end{figure*}

Fig.~\ref{sequence} summarises pedestrian behaviours in relation to pedestrian states and the interrelation between behaviours. Pedestrian states are illustrated with example scenes captured from the videos (Fig.~\ref{state}).

\subsubsection{Movement Adjustment}
\label{adjustment}
We observed a high ratio of movement adjustment in response to the vehicle. The most prominent one is pedestrians adapting their trajectories (n=108, 36\%). This frequently happened when pedestrians were walking towards (n=69, 23\%) or ahead of (n=32, 11\%) the vehicle, potentially causing a conflict of way. Similarly, some pedestrians standing nearby the vehicle made way for it (n=40, 13\%). Some pedestrians stopped and gave way (n=22, 7\%) in situations comparable to crossing scenarios, e.g. walking towards from the side.

We found some group reactions distinct from the behaviour of singletons. The most evident one is that some pedestrians alerted their fellows about the vehicle's presence (n=25, 8\%). This happened when the vehicle was approaching them, and those who noticed the vehicle would notify their friends who were unaware of it. Pedestrians in a group usually acted together when adjusting movements, e.g. all moving to the same side. Only in two interactions (1\%) pedestrians adapted their trajectory at the cost of splitting from their friends.

\subsubsection{Curiosity}
The most common behaviour is curious gazing (n=128, 42\%), which we measured by finding pedestrians turned their head or body at certain angles to stay gazing or gazed multiple times (looked away and back several times). Pedestrians in two interactions (1\%) discussed the vehicle in a group, with smiles or pointing at the vehicle.

\subsubsection{Etiquette}
\label{etiquette}
Pedestrians smiled (n=27, 9\%) or nodded (n=3, 1\%) at the direction of the vehicle\footnote{The reactions could be in response to the vehicle, the driver, or both.} in various situations, for example, along with gazing (n=7, 2\%) or with adapting trajectories (n=7, 2\%). Three pedestrians in three interactions (1\%) conveyed apology (one gestured and two said ``sorry'') in their momentary conflict of passage with the vehicle. Another two pedestrians in two interactions (1\%) gestured to give way.

\begin{figure*}[t]
    \centering
    \begin{subfigure}[b]{0.33\textwidth}
        \centering
        \includegraphics[width=\textwidth]{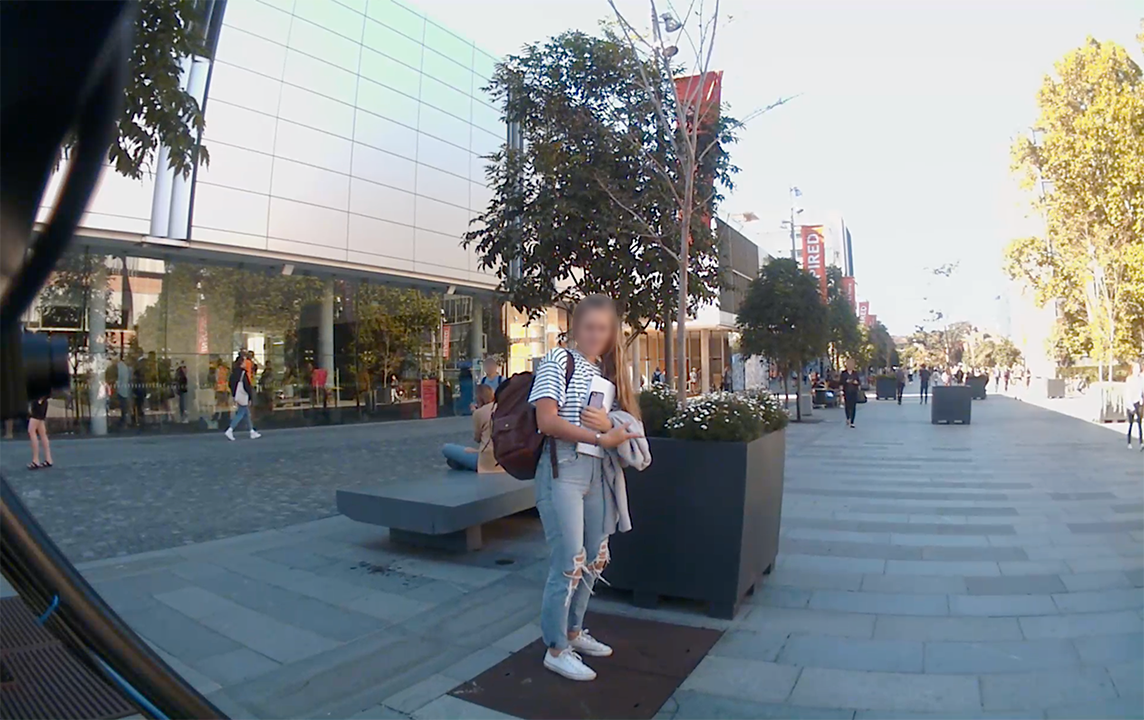}
        \caption{}
        \label{gesture}
    \end{subfigure}
    \begin{subfigure}[b]{0.33\textwidth}
        \centering
        \includegraphics[width=\textwidth]{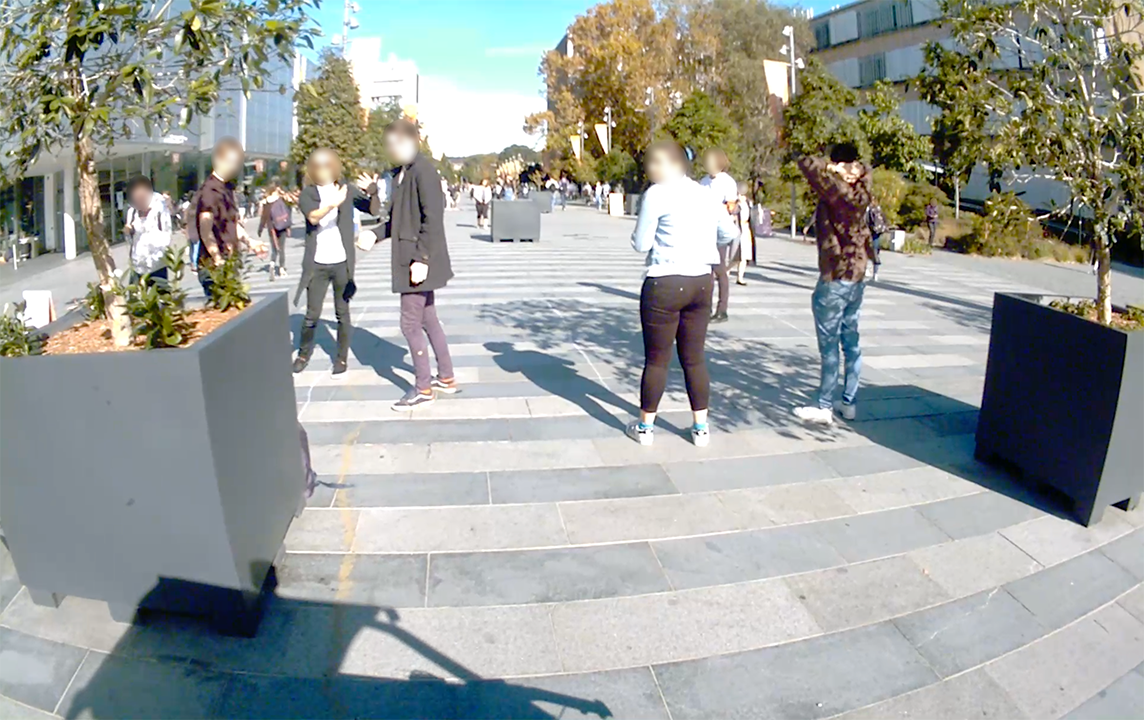}
        \caption{}
        \label{direct1}
    \end{subfigure}
    \begin{subfigure}[b]{0.33\textwidth}
        \centering
        \includegraphics[width=\textwidth]{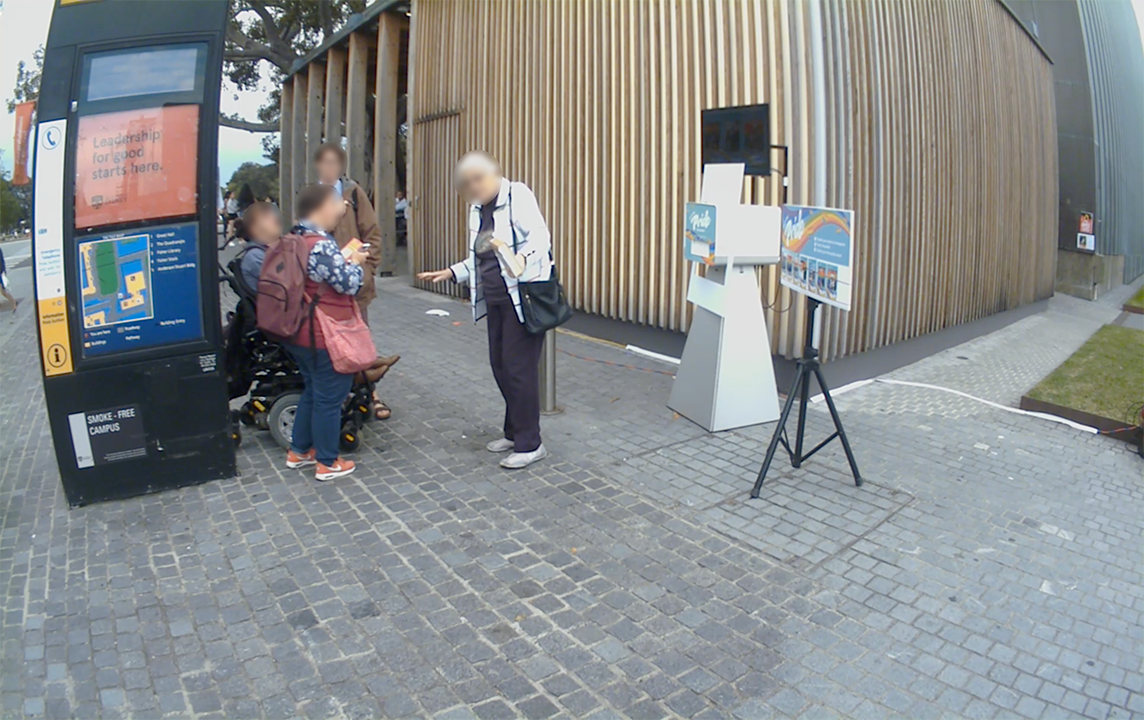}
        \caption{}
        \label{direct2}
    \end{subfigure}
    \caption{(a) The girl was smiling and gesturing to give way to the vehicle. (b) One player in the group was guiding the vehicle to pass through the crowd. (c) An elderly woman was asking the wheelchair user and the two helpers to make way for the vehicle.}
\end{figure*}

\subsubsection{Prosocial Behaviour}
\label{prosocial}
Prosocial behaviour in traffic situations can be defined as actions that benefit other traffic participants, which helps achieve safer and more efficient cooperation and resolve conflicts in a positive manner \cite{harris2014prosocial,sahin2021workshop,sadeghian2020exploration}. Particularly, we identified two situations where some pedestrians voluntarily directed traffic for the vehicle to transit (n=2, 1\%). In one situation (Fig.~\ref{direct1}), a group of six people were playing with a ball, occupying a large space in front of the vehicle. One player noticed the vehicle and made a gesture (crossing his hands before his chest) to indicate to other players to stop playing. He then guided everybody to leave a path and finally gestured for the vehicle to cross through. In the other situation (Fig.~\ref{direct2}), a wheelchair user and two helpers were blocking the narrow entrance of Eastern Avenue, unaware of the vehicle. The vehicle waited for two-and-a-half minutes until an elderly woman passing by noticed the situation. She then communicated with the three people and guided them to vacant the entrance. She watched the vehicle moving through, smiled, and nodded at the driver.

Additionally, as mentioned before, pedestrians actively cooperated their actions, and two gave way explicitly using gestures (e.g. Fig.~\ref{gesture}). Pedestrians in two interactions moved onto the lawn to make more room for the vehicle.

\begin{figure}[H]
    \centering
    \begin{subfigure}[b]{0.95\linewidth}
        \raggedright
        \includegraphics[width=0.95\linewidth]{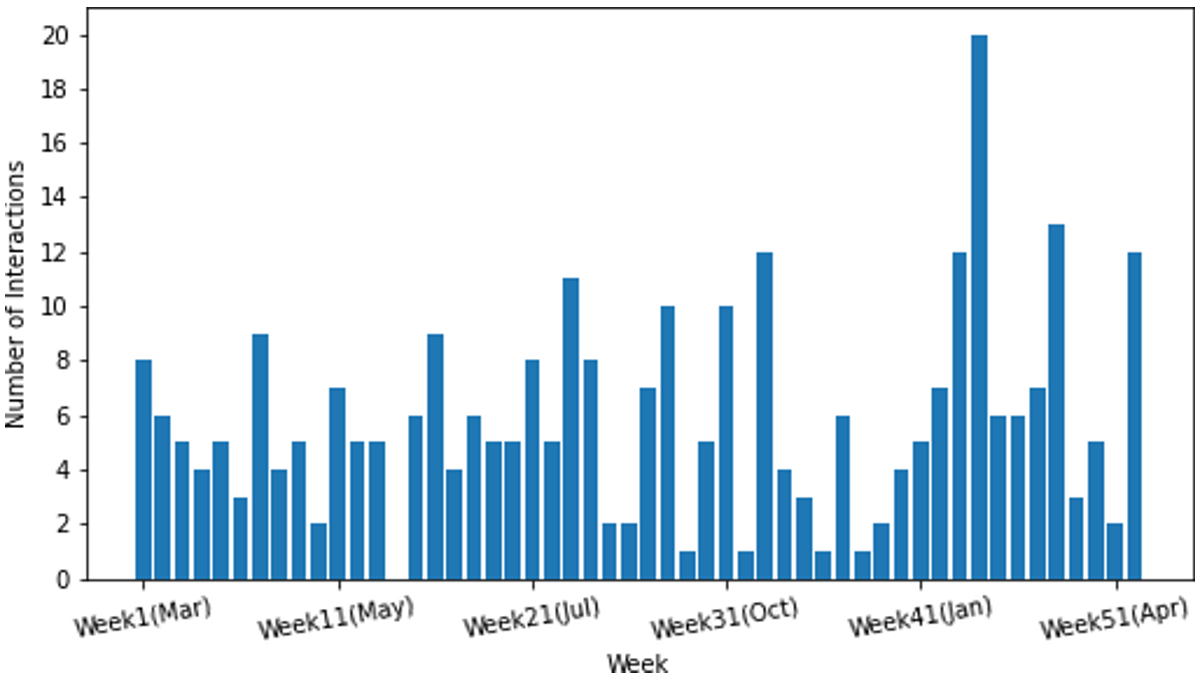}
        \caption{}
        \label{distribution}
    \end{subfigure}
    \begin{subfigure}[b]{0.97\linewidth}
        \centering
        \includegraphics[width=0.97\linewidth]{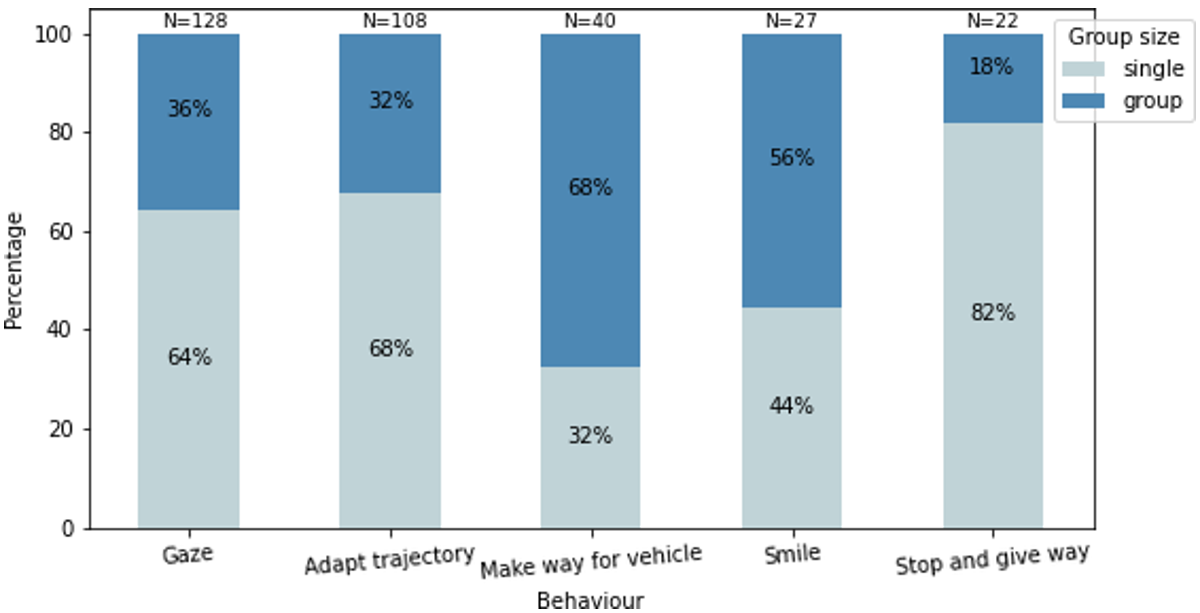}
        \caption{}
        \label{group}
    \end{subfigure}
    \caption{(a) Weekly number of interactions. (b) Percentage of pedestrians in group and alone across five behaviours: \textit{Gaze}, \textit{Adapt trajectory}, \textit{Make way for vehicle}, \textit{Smile}, and \textit{Stop and give way}.}
\end{figure}

\subsubsection{Group Effect}
\label{group_effect}
Fig.~\ref{group} shows the percentage of different group size (single vs. group) in \textit{Gaze}, \textit{Adapt trajectory}, \textit{Make way for vehicle}, \textit{Smile}, and \textit{Stop and give way}. These five behaviours have noticeably greater counts than others (Fig.~\ref{sequence}). We omitted \textit{Alert fellow} because it only existed in groups. We further performed Pearson chi-square tests of association to examine the relationship between group size (single vs. group) and each of the five behaviours. Group size was significantly associated with \textit{Make way for vehicle} ({$\chi^2$}=17.242, df=1, p<0.0005), \textit{Stop and give way} ({$\chi^2$}=3.893, df=1, p=0.049), and \textit{Smile} ({$\chi^2$}=3.959, df=1, p=0.047). Compared to singletons, pedestrians in groups were significantly more likely than expected to make way for the vehicle (Observed = 27, Expected = 15.1) and to smile at the vehicle (Observed = 15, Expected = 10.2), while singletons were significantly more likely than expected to stop and let the vehicle pass first (Observed = 18, Expected = 13.7) compared to pedestrians in groups.

\begin{table*}[t]
  \caption{Critical situations that have safety or efficiency implications.}
  \label{tab:situations}
  \begin{tabular}{p{0.5cm}|p{7cm}|p{7cm}}
    \toprule
    No. & Description of situation (number of interactions) & Observed consequence\\
 \midrule
    1 & Pedestrians were using phones and did not pay attention to the vehicle coming. (29) & Most of them adapted abruptly. Four pedestrians smiled. One pedestrian said ``sorry''.\\[0.5em]
    
    2 & Pedestrians walked hesitantly before crossing ahead of vehicle. (3) & They eventually stopped to let the vehicle pass first. One of them smiled and gestured to give way.\\[0.5em]
    
    3 & Pedestrians were arranging their belongings and did not see the approaching vehicle. (2) & One changed trajectory abruptly; the other slightly adapted to the side and smiled.\\[0.5em]
    
    4 & A young kid suddenly rushed out of the crowd from the left towards the vehicle. (1) & The kid swerved in front of the vehicle and ran to the other side of the road.\\[0.5em]
    
    5 & A young kid was standing ahead, facing the vehicle sideways and being taken a photo by an adult. (1) & The adult quickly helped the kid move to the side after noticing the vehicle.\\[0.5em]
    
    6 & A schoolboy was crossing from the left, gazing elsewhere and not noticing the approaching vehicle. (1) & Notified by another man, the boy was startled and stopped abruptly.\\[0.5em]
    
    7 & A man walking ahead of the vehicle suddenly turned around to walk in the opposite direction. (1) & The man was startled to see the vehicle behind and quickly moved out of the vehicle's way.\\[0.5em]
    
    8 & A man was wandering in front, gazing elsewhere and did not see the approaching vehicle. (1) & The man was surprised, said ``sorry'', smiled, and moved out of the vehicle's way.\\[0.5em]
 
    9 & Three people were standing ahead, facing backwards at the vehicle and blocking a narrow path. (1) & The vehicle slowly rolled forwards and waited for thirty seconds until they noticed it and quickly made way.\\
  \bottomrule
\end{tabular}
\end{table*}

\subsection{Critical Situations}
We observed a number of critical situations from atypical pedestrian reactions during (potential) conflicts of way with the vehicle, which have compromised the safety or efficiency of the interactions (Table~\ref{tab:situations}).

\subsubsection{Inattention}
Pedestrians seemed inattentive to the vehicle in these situations (apart from situation no.2), resulting in risky movement change (e.g. abrupt adjustment) and negative emotions (e.g. fright). Twenty-nine interactions involved pedestrians engaging in their personal devices (no.1). Some children were found to be less vigilant of their surroundings and behave in ways less predictable to the vehicle (no.4-6).

\subsubsection{Emotional Expression}
Most inattentive pedestrians were startled by the approaching vehicle. Some smiled to lighten the situation (no.1,3,8), and some were apologetic. A few pedestrians displayed hesitation when trying to cross ahead of the vehicle (no.2).

\subsection{Additional Observations}
\subsubsection{Communication with Driver}
Pedestrians in four interactions clearly communicated with the vehicle driver. One pedestrian waved a hand to greet the driver. Three pedestrians talked to the driver, with one giving away free bakeries.

\subsubsection{Social Activities}
\label{social_activities}
Apart from traversing the spaces, pedestrians lingered and engaged in various social activities, such as chatting and drinking coffee with friends and taking photos of the surroundings. Occasionally, people set up booths and gave out flyers, and there were a few organised events where vendors came and people gathered.

\section{Discussion}
The presence of the vehicle in the pedestrian areas seems intriguing to many pedestrians as curious gazing was frequently observed. Indeed, vehicular volume was low in those spaces, and its golf cart appearance and signs for automation (cameras, LiDAR, and labels attached to its exterior), might look unusual to the pedestrians. Besides, pedestrian reactions were ordinary to positive, and no hostile behaviours \cite{moore2020defense} were observed. This could be due to the presence of the driver, the car's small, friendly appearance \cite{dey2019pedestrian}, and contextual factors like the university environment and the cultural background \cite{hoggenmueller2020emotional}. In general, the comfortable environment offered by the shared spaces invited a broad range of RUs to stay and make use of the places, which contributed to the breadth of behaviours and situations observed.

Pedestrians used noticeably more implicit norms (movement) than explicit norms (gestures, facial expressions) (Fig.~\ref{sequence}). This is in line with observations on normal roads that pedestrians and vehicles mainly use movement to signal intentions \cite{dey2017pedestrian,risto2017human,moore2019case}. Nevertheless, as shown in Fig.~\ref{sequence}, interactions in shared spaces are no longer confined to the typical orthogonal interactions at crossings, as pedestrians can have varied approaching angles \cite{predhumeau2021pedestrian}, walking directions \cite{madigan2019understanding}, and types of activities (stationary or moving \cite{karndacharuk2013analysis}). In addition, it is interesting to note that information about the vehicle is not only transmitted in the vehicle-pedestrian dyads but also among pedestrians, as we found that multiple pedestrians communicated the information about the vehicle in a group (e.g. \textit{Alert fellow}) or even when they did not know one another (e.g. \textit{Prosocial Behaviour}). Based on the observation, the following sections provide early-staged considerations to support external interaction designs for future AVs in shared spaces. The insights include but go beyond the design of eHMIs as they consider the broader view of interaction in the context of use \cite{beaudouin2004designing}.

\subsection{Movement and Proxemics}
Movement change was the most frequent pedestrian interaction with the vehicle, in that pedestrians constantly adapted trajectory, position, and speed in response to the vehicle. In the context of a shared space, such behaviour can be further associated with a respect for spatial distances from others \cite{predhumeau2021pedestrian}. To that end, we can draw on the notion of proxemics introduced by Hall as a way to describe physical distances in social space as intimate, personal, social, and public distances \cite{hall1968proxemics}. The framework has previously been translated into interaction design techniques that mediate the social encounters between a system and its users \cite{ballendat2010proxemic} and for urban interactive applications \cite{tomitsch2017making}. Similarly, external interactions in shared spaces could consider proxemics to determine the AV's behaviour.

Pedestrian expectations of vehicle proxemics can correlate with their past experience with (but not limited to) vehicles and public spaces. We identified situation no.7 in Table~\ref{tab:situations} -- the pedestrian walking in front of the vehicle appeared startled when he suddenly realised the vehicle was closely behind him. It is then essential to design the use of space for vehicles to follow such tacit norms and avoid eliciting negative emotions in pedestrians. Further, it should be investigated whether and how the expectations would change for different types of vehicles (automation level, size, appearance, etc.). Still, \textbf{future vehicles in shared spaces should adjust movement in relation to proxemics expected by pedestrians.}

\subsection{Propagation of Information}
People in a social environment can share information and behave in an adaptive manner \cite{faria2010collective,gallup2012visual}; for instance, visual attention and emotional state can propagate in human crowds \cite{gallup2012visual,li2020road,barsade2002ripple}, and pedestrians are found to follow others when crossing roads \cite{faria2010collective,hamed2001analysis}. These can link to our analysis of the group effect (Section~\ref{group_effect}), in which we found that pedestrians in groups were more likely to smile at the vehicle or to make way for it.

The propagation of information among pedestrians was observed to benefit the vehicle's movement. Particularly, we extract two scenarios that represent two forms of the information transmission. The first one is derived from \textit{Alert fellow} (n=25) that some pedestrians alerted their fellows of the approaching vehicle, and they moved to avoid collision collectively. In this scenario, pedestrians who gain information directly from the vehicle serve as a ``proxy'' and pass the information on to their acquainted members of a group. The second one is derived from the prosocial situation where one passer-by communicated the vehicle's intention to three nearby people who were unaware that they had been blocking the way for nearly three minutes (Section~\ref{prosocial}). This leads to the scenario where the vehicle fails to communicate with some pedestrians, and others (even though they may not know those pedestrians) act as a ``messenger'' and carry out the communication on the vehicle's behalf.

These observations suggest that pedestrians can gain information not only directly from the vehicle but also from social sources (other pedestrians) \cite{faria2010collective}. It would be interesting to consider how this phenomenon would affect the delivery of information for future AVs, for example, whether the propagation would be more efficient than solely relying on vehicle signals to reach individuals. Besides, the collective response of a group (e.g. avoid the vehicle together) suggests the possibility of treating a group as a unit of interaction subject, expanding the typical one-on-one AV-pedestrian interaction \cite{verstegen2021commdisk}. Additionally, in the case of a voluntary ``messenger'' who assists the AV in communicating its intention, the vehicle can reciprocate the behaviour, such as acknowledging or thanking the messenger \cite{colley2021investigating}. Hence, \textbf{future vehicles in shared spaces should support and acknowledge information propagation among pedestrians.}

\subsection{The Comfortable Environment}
Pedestrians seemed to comfortably roam and use the shared spaces, especially in the two outdoor pedestrian areas (Eastern Avenue and the Quadrangle). This could be due to the invariably low vehicle volume and speed \cite{moody2014shared,kaparias2012analysing} as well as the recreational functionality afforded by the physical design of those spaces \cite{karndacharuk2013analysis}. We frequently found pedestrians traversing the spaces while using their phones or talking with others. Previous studies have found that such technological and social distractions can lead to lower vigilance of surroundings and fewer cautionary behaviours \cite{thompson2013impact,hussein2016p2v}. Indeed, a number of critical situations (Table~\ref{tab:situations}) were linked to pedestrian inattention, many of which were due to phone use. These situations compromised the safety or efficiency of the interactions and often caused unpleasant experiences for pedestrians (e.g. negative emotions). It is therefore important to design appropriate notification mechanisms that raise pedestrians' awareness of nearby vehicles, particularly for those who seem inattentive to their surroundings \cite{hussein2016p2v}.

Vehicle-to-pedestrian warnings or notifications should be carefully designed for a shared space context. Place-making is amongst the primary objectives of the shared space approach, which aims to improve the ambience of the streets and the well-being of pedestrians \cite{moody2014shared,karndacharuk2013analysis}. We have observed various activities and events where people gathered and enjoyed their time (Section~\ref{social_activities}). Hence, less obtrusive mechanisms should be considered (e.g. a low decibel, synthetic motor sound \cite{moore2020sound}) since common attention-grabbing warnings (e.g. a traditional horn or a flashing eHMI \cite{li2018cross}) can potentially disrupt the atmosphere despite that they can raise immediate awareness. Thus, \textbf{future vehicles in shared spaces should balance raising pedestrians' awareness and sustaining the ambience of the spaces.}

\subsection{Limitations and Future Work}
We acknowledge that there are some limitations to this study. Firstly, the presence of a driver could have influenced pedestrian behaviours. For example, though we reported the importance of movement and proxemics, specific requirements (e.g. time and distance) could differ for fully AVs \cite{hoggenmueller2022designing}. As such, this study provides only a first foundation-setting step into the current pedestrian-vehicle interactions in shared spaces. The qualitative results facilitate understanding towards the possible, various forms of interaction extracted from the unstructured and ambiguous scenarios.

Secondly, although the behaviours are observed from a long-term base which potentially reduces the novelty effect over time, the context of the observation is confined by space (the university campus). It seems to us that pedestrians were mostly students and staff from the university and thus their behaviours might not represent a larger population. Besides, the peaceful campus could have influenced people's mood and their speculations about the vehicle's purpose in a positive way \cite{hoggenmueller2020emotional}. Nonetheless, the study featured two outdoor shared spaces with a high pedestrian focus and recreational functionalities, which could resemble or have implications for urban public spaces sharing the similar motivations behind the deployment. The small electric vehicle used in this study -- though it might not represent heavier vehicles normally seen on motorways -- had the advantage of resembling a hybrid of vehicles and mobile robots (e.g. self-driving pod cars, delivery vehicles) that are likely to roam such urban shared spaces in the near future.

\section{Conclusion}
We have observed a range of pedestrian-vehicle interactions from the 14-month dataset collected in shared spaces. Movement was the most frequent form of pedestrian interaction with the vehicle, which is also associated with norms of proxemics used by people when sharing public spaces. Pedestrians in crowds led to group behaviours different from singletons, and interestingly, the propagation of information among pedestrians even in situations where they did not know each other. Moreover, pedestrians demonstrated prosocial behaviours that benefited the vehicle's passage. Nevertheless, the comfortable environment offered by the shared spaces might result in pedestrians lowering the vigilance of their surroundings, and vehicle-to-pedestrian warning mechanisms should be appropriately designed in these situations. Drawn from current real-world pedestrian interactions with the small, manually-driven vehicle in shared spaces, our findings present both challenges and opportunities unique to the shared space context at an early, exploratory stage and serve as a foundation for future work that supports AVs operating in such environments. The insights include but go beyond designing interfaces attached to the exterior of AVs and prompt a broader view of interaction in the context of use, which is especially critical in shared spaces that afford a variety of pedestrian states and behaviours.

\begin{acks}
This study was funded by the Australian Research Council through grant number DP200102604 Trust and Safety in Autonomous Mobility Systems: A Human-Centered Approach. We thank the Intelligent Transport Systems Research Lab at the Australian Centre for Field Robotics for collecting the data. We thank our colleagues in the Design Lab, Marius Hoggenmueller and Tram Thi Minh Tran, for their valuable feedback and continuous support. We thank Kathrin Schemann in the Sydney Informatics Hub at the University of Sydney for providing statistical consultancy. This study has been approved by the Human Research Ethics Committee at the University of Sydney with protocol number 2022/193.
\end{acks}

\bibliographystyle{ACM-Reference-Format}
\bibliography{reference}


\end{document}